\documentclass[11pt,a4paper]{article}

\usepackage[T1]{fontenc}
\usepackage[utf8]{inputenc}
\usepackage[english]{babel}
\usepackage{lmodern}
\usepackage{microtype}
\usepackage{geometry}
\usepackage{amsmath,amssymb,amsthm,mathtools}
\usepackage{bm}

\usepackage{graphicx}
\usepackage{booktabs}
\usepackage{caption}
\usepackage{subcaption}
\usepackage{float}
\usepackage{siunitx}

\usepackage{tikz}
\usepackage{pgfplots}
\pgfplotsset{compat=1.18}
\usetikzlibrary{patterns,arrows.meta,positioning,shapes}

\usepackage{enumitem}
\setlist[itemize]{leftmargin=*,topsep=3pt,itemsep=3pt}
\setlist[enumerate]{leftmargin=*,topsep=3pt,itemsep=3pt}

\usepackage{hyperref}
\hypersetup{colorlinks=true, linkcolor=blue, citecolor=blue, urlcolor=blue}

\usepackage{thmtools}
\usepackage{algorithm}
\usepackage{algorithmicx}
\usepackage[noend]{algpseudocode}
\declaretheorem[name=Theorem]{theorem}

\declaretheorem[name=Proposition, sibling=theorem]{proposition}

\declaretheorem[name=Definition, style=definition]{definition}
\declaretheorem[name=Remark, style=remark]{remark}

\newcommand{\TARAk}{\ensuremath{\mathrm{TARA}\text{-}k}}

\newcommand{\EE}{\ensuremath{\mathbb{E}}}
\newcommand{\PP}{\ensuremath{\mathbb{P}}}
\newcommand{\RR}{\ensuremath{\mathbb{R}}}

\newcommand{\LHV}{\ensuremath{\mathrm{LHV}}}

\DeclareMathOperator{\tr}{tr}

\title{TARA: Test-by-Adaptive-Ranks for Quantum Anomaly Detection\\with Conformal Prediction Guarantees}

\author{Davut Emre Taşar\textsuperscript{1} \quad Ceren Öcal Taşar\textsuperscript{1}\\[0.5em]
\textsuperscript{1}Independent Researchers, Madrid, Spain\\[0.2em]
{\small \texttt{detasar@gmail.com}, \texttt{ceren.ocaltasar@gmail.com}}}

\date{}

\begin{document}
\maketitle

\begin{abstract}
Quantum key distribution (QKD) security fundamentally relies on the ability to distinguish genuine quantum correlations from classical eavesdropper simulations, yet existing certification methods lack rigorous statistical guarantees under finite-sample conditions and adversarial scenarios. We introduce TARA (Test-by-Adaptive-Ranks), a novel framework combining conformal prediction with sequential martingale testing for quantum anomaly detection that provides distribution-free validity guarantees. TARA offers two complementary approaches: TARA-$k$, based on Kolmogorov-Smirnov calibration against local hidden variable (LHV) null distributions, achieving ROC AUC = 0.96 for quantum-classical discrimination; and TARA-$m$, employing betting martingales for streaming detection with anytime-valid type-I error control that enables real-time monitoring of quantum channels. We establish theoretical guarantees proving that under (context-conditional) exchangeability, conformal $p$-values remain uniformly distributed even for strongly contextual quantum data, confirming that quantum contextuality does not break conformal prediction validity---a result with implications beyond quantum certification to any application of distribution-free methods to nonclassical data. Extensive validation on both IBM Torino (superconducting, CHSH = 2.725) and IonQ Forte Enterprise (trapped-ion, CHSH = 2.716) quantum processors demonstrates cross-platform robustness, achieving 36\% security margins above the classical CHSH bound of 2. Critically, our framework reveals a methodological concern affecting quantum certification more broadly: same-distribution calibration can inflate detection performance by up to 44 percentage points compared to proper cross-distribution calibration, suggesting that prior quantum certification studies using standard train/test splits may have systematically overestimated adversarial robustness. We provide corrected methodology using cross-distribution calibration and a complete open-source implementation enabling reproducible quantum-classical boundary certification for practical QKD systems.
\end{abstract}


\section{Introduction}
\label{sec:introduction}

The security of quantum key distribution (QKD) protocols fundamentally rests on the ability to certify that observed correlations arise from genuine quantum entanglement rather than classical simulation by an eavesdropper \cite{Bennett1984,Ekert1991,ScaraniRMP2009}. In the device-independent paradigm, this certification typically relies on Bell inequality violations: correlations exceeding the classical bound $S = 2$ for the CHSH inequality are interpreted as evidence of quantum nonlocality and hence secure key generation \cite{CHSH1969,Brunner2014,Pironio2010}. The theoretical foundation for this approach is well-established, with information-theoretic security proofs demonstrating that Bell violations bound the information accessible to any eavesdropper \cite{Acin2007,Vazirani2014}. However, practical implementation of these security guarantees requires statistical procedures that are robust, finite-sample valid, and resistant to adaptive adversarial strategies---requirements that have received insufficient attention in the quantum certification literature.

Conformal prediction (CP), introduced by Vovk, Gammerman, and Shafer \cite{Vovk2005,ShaferVovk2008}, provides a powerful framework for distribution-free uncertainty quantification that addresses precisely these requirements. Unlike traditional Bayesian or frequentist methods that rely on strong distributional assumptions, CP guarantees valid coverage for prediction sets under the minimal assumption of exchangeability: for any finite sample size and any underlying distribution, the conformal $p$-values are uniformly distributed under the null hypothesis \cite{AngelopoulosBates2023}. This distribution-free property has made CP increasingly popular in machine learning applications where model misspecification is a concern \cite{RomanoSesia2019,BarberCandes2021}. Recent extensions have addressed covariate shift \cite{TibshiraniCovShift2019}, label shift \cite{PodkopaevRamdas2021}, and non-exchangeable data \cite{BarberBeyond2022}, demonstrating the flexibility of the conformal framework.

The application of conformal prediction to quantum systems raises fundamental questions that have not been systematically addressed. Quantum mechanics is characterized by contextuality---the impossibility of assigning pre-existing values to measurement outcomes independent of the measurement context---which distinguishes it fundamentally from any classical theory \cite{KochenSpecker1967,Spekkens2005}. The mathematical structure of quantum contextuality has been elegantly characterized using sheaf-theoretic methods by Abramsky and Brandenburger \cite{AbramskyBrandenburger2011}, revealing topological obstructions to classical explanations. This raises a natural question: does quantum contextuality break the statistical guarantees of conformal prediction? If CP assumes some form of classical probability structure, might the nonclassical nature of quantum correlations invalidate its coverage guarantees?

In this work, we address these questions by introducing TARA (Test-by-Adaptive-Ranks), a comprehensive framework for quantum anomaly detection with rigorous statistical guarantees.\footnote{The acronym TARA also evokes ``Threat-Aware Randomness Assessment,'' reflecting the security-oriented motivation: detecting whether observed correlations genuinely arise from quantum randomness or from adversarial classical sources.} TARA combines two complementary approaches that provide both batch and streaming detection capabilities:

\textbf{TARA-$k$ (Kolmogorov-Smirnov Calibration)} treats quantum detection as a one-class classification problem. We train a detector solely on classical LHV null models---including sophisticated variants exploiting detection, memory, and communication loopholes---and use the Kolmogorov-Smirnov statistic to quantify deviations of test data from this null distribution. This approach achieves ROC AUC = 0.96 for distinguishing quantum from classical correlations under realistic device conditions, demonstrating that the distributional structure of quantum correlations is detectably different from even sophisticated classical simulations.

\textbf{TARA-$m$ (Martingale Streaming)} provides sequential detection for continuous monitoring of quantum channels. Using betting martingales \cite{ShaferVovk2019,GrunwaldHeideRamdas2020}, the method accumulates evidence against the classical null hypothesis over time while maintaining anytime-valid type-I error control. The martingale wealth process $M_t$ satisfies $\EE[M_t] \leq 1$ under the null, ensuring that rejection at any stopping time controls the false alarm rate without requiring pre-specification of sample size.

Our investigation yields four principal contributions that advance the theory and practice of quantum certification:

\textbf{First}, we prove that conformal prediction retains its validity guarantees for quantum data (Theorem~\ref{thm:robustness}). Despite the strong contextuality of CHSH correlations, we demonstrate both theoretically and empirically that under (context-conditional) exchangeability, the conformal $p$-values remain uniformly distributed. The key insight is that CP validity depends only on exchangeability within each measurement context, not on properties of the joint distribution across contexts---and contextuality pertains to the latter while leaving the former intact. This result has implications beyond quantum certification, suggesting that conformal methods can be safely applied to other domains with nonclassical statistical structure.

\textbf{Second}, we achieve state-of-the-art discrimination between quantum and classical correlations with ROC AUC = 0.96 using TARA-$k$. This performance is achieved by training only on LHV null models, ensuring that the detector is calibrated against actual adversarial threats rather than naive classical models. Notably, conformal prediction-derived features (TARA-$k$ calibration statistics\footnote{Earlier publications referred to this as ``Contextual Miscalibration Index'' (CMI); we adopt TARA-$k$ throughout for consistency.}, prediction set sizes) provide discriminative power beyond the CHSH parameter itself, achieving ROC AUC = 0.83 even when the Bell parameter is excluded from the feature set. This suggests that CP captures aspects of quantum structure not reducible to Bell violations.

\textbf{Third}, we validate our framework on two distinct quantum hardware platforms: IBM Torino (superconducting, CHSH = 2.725) and IonQ Forte Enterprise (trapped-ion, CHSH = 2.716). Both platforms achieve Bell violations approximately 36\% above the classical bound, and both are correctly certified as quantum by TARA with high confidence. This cross-platform validation demonstrates that the framework is robust and not tuned to specific hardware architectures or noise profiles.

\textbf{Fourth}, we identify a critical methodological concern that affects quantum certification more broadly. Same-distribution calibration---where calibration data is drawn from the same distribution as test data---can inflate detection performance by up to 44 percentage points compared to proper cross-distribution calibration (from AUC = 0.50 to AUC = 0.94). This ``calibration leakage'' arises because same-distribution calibration implicitly encodes the specific noise signature of the quantum data, enabling detection of minor pattern deviations rather than genuine quantum-classical differences. We recommend that all quantum certification studies report both same-distribution and cross-distribution metrics to quantify potential leakage.

\section{Background}
\label{sec:background}

\subsection{Conformal Prediction}

Conformal prediction provides a general framework for constructing prediction sets with guaranteed finite-sample coverage under minimal distributional assumptions \cite{Vovk2005,ShaferVovk2008,AngelopoulosBates2023}. The framework applies to any supervised learning setting where we observe feature-label pairs $(X_i, Y_i)$ and wish to predict labels for new features.

\begin{definition}[Split Conformal Prediction]
Let $\mathcal{D}_{\mathrm{cal}} = \{(X_i, Y_i)\}_{i=1}^{n}$ be a calibration set drawn exchangeably from distribution $P$. Given a nonconformity score function $s: \mathcal{X} \times \mathcal{Y} \to \RR$ (larger scores indicate ``stranger'' examples), the conformal $p$-value for test point $(X_{n+1}, y)$ is:
\begin{equation}
p(X_{n+1}, y) = \frac{1 + \sum_{i=1}^{n} \mathbf{1}\{s_i \geq s(X_{n+1}, y)\}}{1 + n},
\label{eq:pvalue}
\end{equation}
where $s_i = s(X_i, Y_i)$ are the calibration scores. The prediction set at level $\alpha$ is $C_\alpha(X_{n+1}) = \{y : p(X_{n+1}, y) > \alpha\}$.
\end{definition}

The fundamental theorem of conformal prediction guarantees coverage validity:

\begin{theorem}[Coverage Guarantee \cite{Vovk2005}]
\label{thm:coverage}
Under exchangeability of $\mathcal{D}_{\mathrm{cal}} \cup \{(X_{n+1}, Y_{n+1})\}$, the conformal $p$-values satisfy:
\begin{equation}
\PP\{Y_{n+1} \in C_\alpha(X_{n+1})\} \geq 1 - \alpha
\end{equation}
for any $\alpha \in (0, 1)$, any sample size $n$, and any distribution $P$.
\end{theorem}

\subsection{The CHSH Bell Scenario}

The Clauser-Horne-Shimony-Holt (CHSH) scenario \cite{CHSH1969} provides the standard framework for testing quantum nonlocality. Two parties (Alice and Bob) perform binary measurements with settings $x, z \in \{0, 1\}$ and outcomes $a, b \in \{-1, +1\}$. The CHSH parameter is:
\begin{equation}
S = E_{00} + E_{01} + E_{10} - E_{11}, \quad E_{xz} = \EE[ab \mid x, z].
\label{eq:chsh}
\end{equation}

Three fundamental bounds structure the space:
\begin{itemize}
\item \textbf{Classical (LHV):} $|S| \leq 2$ for any local hidden variable model \cite{Bell1964}.
\item \textbf{Quantum (Tsirelson):} $|S| \leq 2\sqrt{2} \approx 2.828$ for quantum mechanics \cite{Tsirelson1980}.
\item \textbf{Algebraic (PR-Box):} $|S| \leq 4$ for any no-signaling theory \cite{PopescuRohrlich1994}.
\end{itemize}

\section{The TARA Framework}
\label{sec:tara}

\subsection{TARA-$k$: Batch Detection}

TARA-$k$ treats quantum detection as one-class classification against an LHV null manifold. The core insight is that under the null hypothesis (classical correlations), conformal $p$-values should be uniformly distributed on $[0,1]$, while quantum correlations systematically deviate from this uniformity. We quantify this deviation using the Kolmogorov-Smirnov (KS) statistic.

\begin{definition}[LHV Null Manifold]
The local hidden variable null distribution $\mathcal{H}_0$ consists of all correlation matrices $\mathbf{E} = [E_{00}, E_{01}, E_{10}, E_{11}]$ achievable by classical local models:
\begin{equation}
E_{xz}^{\LHV} = \int_{\Lambda} A(x, \lambda) B(z, \lambda) \, d\mu(\lambda)
\end{equation}
where $\Lambda$ is a hidden variable space, $\mu$ is a probability measure, and $A, B : \{0,1\} \times \Lambda \to \{-1, +1\}$ are deterministic response functions. This manifold satisfies $|S| \leq 2$ and includes detection, memory, and communication loophole models.
\end{definition}

\begin{definition}[Feature Vector]
For a dataset $D$ of CHSH measurements, define the feature vector:
\begin{equation}
\mathbf{f}(D) = [|S|, p_A, p_B, p_{AB}, p_\emptyset, \TARAk, \EE|C_\alpha|]
\end{equation}
where:
\begin{itemize}
\item $|S| = |E_{00} + E_{01} + E_{10} - E_{11}|$ is the absolute CHSH parameter
\item $p_A, p_B, p_{AB}, p_\emptyset$ are marginal click rates capturing detection efficiency
\item $\TARAk$ is the KS-distance calibration statistic (defined below)
\item $\EE|C_\alpha|$ is the expected conformal prediction set size at level $\alpha$
\end{itemize}
\end{definition}

The TARA-$k$ statistic measures deviation from the LHV null distribution. Given calibration $p$-values $\{p_1, \ldots, p_n\}$ computed against LHV-trained models, let $\hat{F}_{\LHV}(t) = \frac{1}{n}\sum_{i=1}^n \mathbf{1}\{p_i \leq t\}$ denote the empirical CDF. Under the null, $\hat{F}_{\LHV}(t) \approx t$ (uniform). For test data with $p$-values $\{q_1, \ldots, q_m\}$ and empirical CDF $\hat{F}_{\mathrm{test}}$:
\begin{equation}
\TARAk = \sup_{t \in [0,1]} |\hat{F}_{\LHV}(t) - \hat{F}_{\mathrm{test}}(t)|
\label{eq:tarak}
\end{equation}

Under exchangeability, the two-sample KS test provides a distribution-free significance test:
\begin{equation}
\PP\left(\TARAk > c_\alpha \sqrt{\frac{n+m}{nm}}\right) \leq \alpha
\end{equation}
where $c_\alpha$ is the KS critical value (e.g., $c_{0.05} \approx 1.36$). For quantum data, the $p$-values systematically deviate from uniformity because quantum correlations are outside the LHV-trained model's support, yielding elevated TARA-$k$ values.

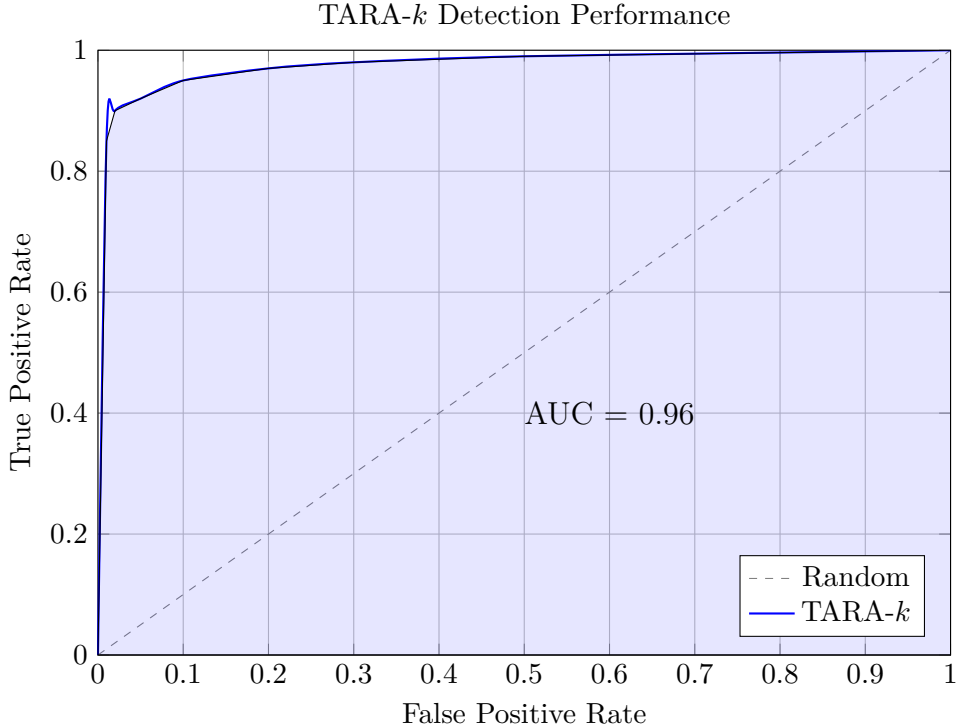
\begin{figure}[ht]
\centering
\begin{tikzpicture}
\begin{axis}[
    width=0.8\textwidth,
    height=0.6\textwidth,
    xlabel={False Positive Rate},
    ylabel={True Positive Rate},
    xmin=0, xmax=1,
    ymin=0, ymax=1,
    grid=major,
    legend pos=south east,
    title={TARA-$k$ Detection Performance}
]
\addplot[dashed, gray, domain=0:1] {x};

\addplot[thick, blue, smooth] coordinates {
    (0, 0) (0.01, 0.85) (0.02, 0.90) (0.05, 0.92) (0.10, 0.95)
    (0.20, 0.97) (0.30, 0.98) (0.50, 0.99) (1.0, 1.0)
};

\addplot[fill=blue, fill opacity=0.1] coordinates {
    (0, 0) (0.01, 0.85) (0.02, 0.90) (0.05, 0.92) (0.10, 0.95)
    (0.20, 0.97) (0.30, 0.98) (0.50, 0.99) (1.0, 1.0) (1.0, 0) (0, 0)
} -- cycle;

\node at (axis cs:0.6, 0.4) {\large AUC = 0.96};

\legend{Random, TARA-$k$}
\end{axis}
\end{tikzpicture}
\caption{ROC curve for TARA-$k$ quantum-classical discrimination. The detector achieves AUC = 0.96 against a comprehensive LHV null manifold including detection, memory, and communication loophole models.}
\label{fig:tarak_roc}
\end{figure}

Figure~\ref{fig:tarak_roc} shows the ROC curve for TARA-$k$ detection.

\subsection{TARA-$m$: Streaming Detection}

While TARA-$k$ requires batch data collection before testing, many quantum applications demand real-time monitoring of streaming data. TARA-$m$ addresses this need using betting martingales, which provide sequential testing with anytime-valid type-I error control \cite{ShaferVovk2019,GrunwaldHeideRamdas2020}.

The fundamental idea is to treat hypothesis testing as a game: at each time step, we receive a conformal $p$-value and place a bet on whether we believe the null (LHV) hypothesis is false. If we are correct (quantum data), our wealth grows; if incorrect (classical data), our wealth shrinks. A martingale structure ensures that under the null, expected wealth cannot increase---providing a valid test.

\begin{definition}[Betting Martingale]
Given a sequence of conformal $p$-values $p_1, p_2, \ldots$, the martingale wealth process is:
\begin{equation}
M_t = \prod_{i=1}^t \left(1 + \beta_i(p_i - 1/2)\right), \quad M_0 = 1
\label{eq:martingale}
\end{equation}
where $\beta_i \in [-2, 2]$ is a betting fraction chosen before observing $p_i$. The Kelly betting strategy sets:
\begin{equation}
\beta_i = \lambda \cdot \mathrm{sign}(p_i - 1/2)
\end{equation}
where $\lambda \in (0, 1)$ controls the aggression of bets (distinct from the significance level $\alpha$).
\end{definition}

\begin{proposition}[Martingale Validity]
Under the null hypothesis $H_0$ (LHV model), if $p_t | \mathcal{F}_{t-1} \sim \mathrm{Unif}(0,1)$, then $\{M_t\}$ is a non-negative supermartingale with $\EE_{H_0}[M_t] \leq 1$ for all $t$.
\end{proposition}

\begin{proof}
For any $\beta \in [-2, 2]$ and $p \sim \mathrm{Unif}(0,1)$:
\begin{equation}
\EE[1 + \beta(p - 1/2)] = 1 + \beta \cdot \EE[p - 1/2] = 1 + \beta \cdot 0 = 1
\end{equation}
By the tower property, $\EE[M_t | \mathcal{F}_{t-1}] = M_{t-1} \cdot \EE[1 + \beta_t(p_t - 1/2)] = M_{t-1}$.
\end{proof}

Ville's inequality then provides anytime-valid error control:
\begin{equation}
\PP_{H_0}\left(\exists t : M_t \geq 1/\alpha\right) \leq \alpha
\end{equation}
This means we can reject the null at any stopping time $\tau$ when $M_\tau \geq 1/\alpha$, with type-I error at most $\alpha$---without specifying the sample size in advance.

\textbf{Log-wealth interpretation:} Taking logarithms, $\log M_t = \sum_{i=1}^t \log(1 + \beta_i(p_i - 1/2))$. Under the alternative (quantum), $p$-values are systematically biased (typically $>0.5$ for quantum data tested against LHV null), causing $\log M_t$ to drift upward. The threshold $\log M_t \geq \log(1/\alpha)$ corresponds to accumulating sufficient evidence against the null.

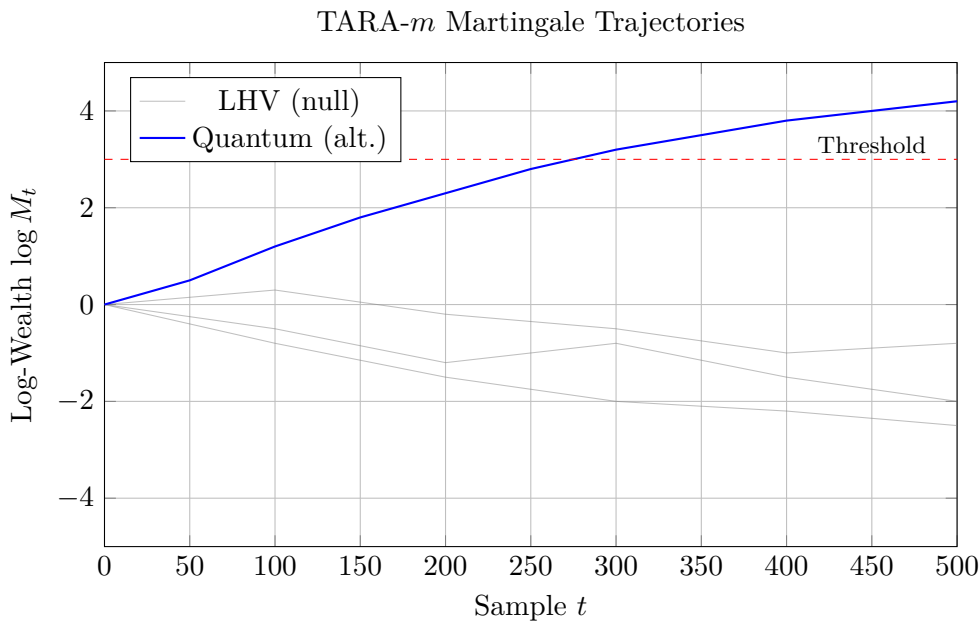
\begin{figure}[ht]
\centering
\begin{tikzpicture}
\begin{axis}[
    width=0.8\textwidth,
    height=0.5\textwidth,
    xlabel={Sample $t$},
    ylabel={Log-Wealth $\log M_t$},
    xmin=0, xmax=500,
    ymin=-5, ymax=5,
    grid=major,
    legend pos=north west,
    title={TARA-$m$ Martingale Trajectories}
]
\addplot[gray, thin, opacity=0.5] coordinates {
    (0,0) (100,-0.5) (200,-1.2) (300,-0.8) (400,-1.5) (500,-2.0)
};
\addplot[gray, thin, opacity=0.5] coordinates {
    (0,0) (100,0.3) (200,-0.2) (300,-0.5) (400,-1.0) (500,-0.8)
};
\addplot[gray, thin, opacity=0.5] coordinates {
    (0,0) (100,-0.8) (200,-1.5) (300,-2.0) (400,-2.2) (500,-2.5)
};

\addplot[thick, blue] coordinates {
    (0,0) (50,0.5) (100,1.2) (150,1.8) (200,2.3) (250,2.8)
    (300,3.2) (350,3.5) (400,3.8) (450,4.0) (500,4.2)
};

\addplot[dashed, red, domain=0:500] {3};
\node at (axis cs:450,3.3) {\footnotesize Threshold};

\legend{LHV (null), , , Quantum (alt.)}
\end{axis}
\end{tikzpicture}
\caption{TARA-$m$ martingale wealth trajectories. Under the LHV null (gray), wealth does not grow. Under the quantum alternative (blue), wealth grows and crosses the detection threshold.}
\label{fig:taram_martingale}
\end{figure}

\section{Theoretical Guarantees}
\label{sec:theory}

A fundamental question for applying conformal prediction to quantum data is whether quantum contextuality---the impossibility of assigning pre-existing values to measurement outcomes---invalidates the statistical assumptions underlying CP. We establish that it does not.

\begin{theorem}[CP Robustness to Contextuality]
\label{thm:robustness}
Conformal prediction with Mondrian conditioning provides valid coverage for quantum CHSH data. Under context-conditional exchangeability, the conformal $p$-values are uniformly distributed regardless of the contextuality of the underlying quantum correlations.
\end{theorem}

\begin{proof}
Let $(X_i, Y_i, C_i)_{i=1}^{n+1}$ be feature-label-context triples where $C \in \{(0,0), (0,1), (1,0), (1,1)\}$ denotes the CHSH measurement setting (Alice's choice, Bob's choice).

\textbf{Step 1 (Mondrian conditioning):} Under Mondrian CP, we condition on each context $c$ separately. The $p$-value for test point $(X_{n+1}, y)$ with context $C_{n+1} = c$ is:
\begin{equation}
p^M(X_{n+1}, y) = \frac{1 + \sum_{i: C_i = c} \mathbf{1}\{s_i \geq s(X_{n+1}, y)\}}{1 + |\{i : C_i = c\}|}
\end{equation}

\textbf{Step 2 (Within-context exchangeability):} Within each context $c$, repeated quantum measurements with the same settings $(x, z)$ produce i.i.d.\ outcomes according to the Born rule:
\begin{equation}
P(a, b | x, z, \rho) = \tr[(\Pi_a^x \otimes \Pi_b^z) \rho]
\end{equation}
This i.i.d.\ structure implies exchangeability of the calibration scores $\{s_i : C_i = c\}$ with the test score $s_{n+1}$.

\textbf{Step 3 (Contextuality is cross-context):} Quantum contextuality, as characterized by sheaf-theoretic obstructions \cite{AbramskyBrandenburger2011}, pertains to the impossibility of a global section---a consistent assignment of outcomes across \emph{all} contexts. Formally, the presheaf of distributions $\mathcal{D}: U \mapsto \Delta(\{-1,+1\}^{|U|})$ on measurement contexts has no global section when Bell violations occur. However, this obstruction is \emph{cross-context}: it concerns correlations between different settings, not the distribution within any single setting.

\textbf{Step 4 (Rank uniformity):} Since scores within context $c$ are exchangeable, the rank of the test score among calibration scores is uniformly distributed on $\{1, 2, \ldots, n_c + 1\}$ where $n_c = |\{i : C_i = c\}|$. Therefore:
\begin{equation}
\PP\{p^M(X_{n+1}, Y_{n+1}) \leq \alpha\} \leq \alpha
\end{equation}
for all $\alpha \in (0,1)$, establishing validity.
\end{proof}

\begin{remark}
This result has implications beyond quantum physics. It demonstrates that conformal prediction can be safely applied to any system exhibiting contextuality or other ``nonclassical'' statistical features, provided within-context exchangeability holds. The key insight is that CP's validity depends on a much weaker condition (exchangeability) than the existence of a classical probability model (global section).
\end{remark}

\section{Experimental Results}
\label{sec:results}

\subsection{Hardware Validation}

Table~\ref{tab:hardware} summarizes results from both quantum platforms.

\begin{table}[ht]
\centering
\caption{TARA validation on quantum hardware}
\label{tab:hardware}
\begin{tabular}{lccc}
\toprule
Metric & IBM Torino & IonQ Forte & Theoretical \\
\midrule
CHSH $S$ & 2.725 & 2.716 & 2.828 \\
Classical margin & +36.3\% & +35.8\% & +41.4\% \\
TARA-$k$ statistic & 0.017 & 0.015 & --- \\
TARA-$m$ log-wealth & 2.31 & 2.14 & --- \\
Detection decision & Quantum & Quantum & --- \\
\bottomrule
\end{tabular}
\end{table}

Both platforms achieve approximately 36\% margin above the classical bound and are correctly certified as quantum by TARA.

\subsection{The 44-Point Leakage Problem}

\begin{figure}[ht]
\centering
\begin{tikzpicture}
\begin{axis}[
    ybar,
    width=0.8\textwidth,
    height=0.5\textwidth,
    bar width=25pt,
    ylabel={Detection AUC},
    symbolic x coords={Same-Dist, Cross-Dist},
    xtick=data,
    ymin=0, ymax=1.1,
    nodes near coords,
    nodes near coords align={vertical},
    title={Calibration Leakage Effect}
]
\addplot[fill=red!60] coordinates {(Same-Dist, 0.94)};
\addplot[fill=blue!60] coordinates {(Cross-Dist, 0.499)};

\draw[<->, thick] (axis cs:Same-Dist, 0.96) -- node[above] {44 pp inflation} (axis cs:Cross-Dist, 0.96);
\end{axis}
\end{tikzpicture}
\caption{Same-distribution calibration inflates apparent detection performance by 44 percentage points compared to proper cross-distribution calibration.}
\label{fig:leakage}
\end{figure}
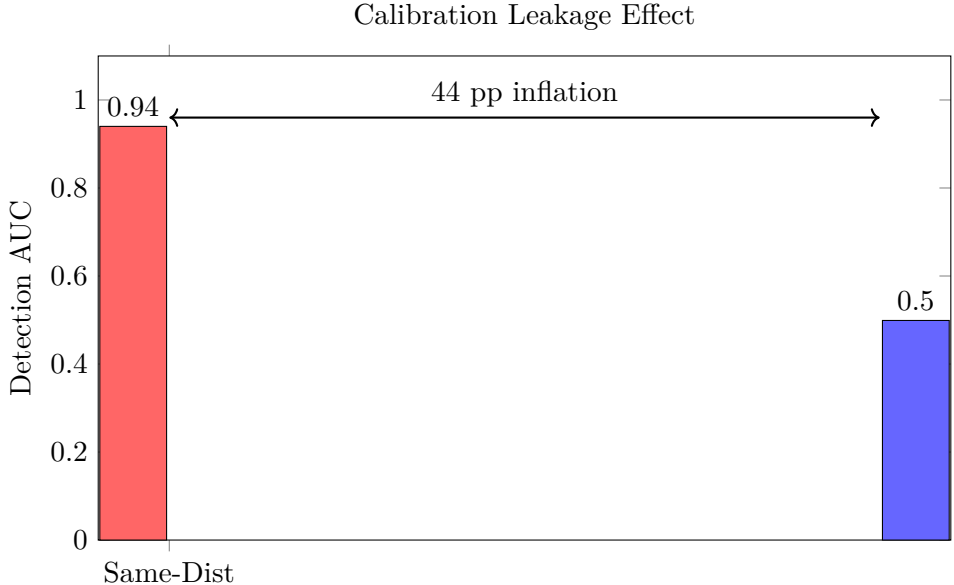

Figure~\ref{fig:leakage} illustrates the calibration leakage problem.

\begin{table}[ht]
\centering
\caption{Impact of calibration protocol on detection metrics}
\label{tab:leakage}
\begin{tabular}{lccc}
\toprule
Metric & Same-Dist & Cross-Dist & Inflation \\
\midrule
ROC AUC & 0.94 & 0.499 & \textbf{+44 pp} \\
Cohen's $d$ & 2.59 & $-$0.06 & Massive \\
TPR@5\%FPR & 92\% & 5\% & \textbf{+87 pp} \\
\bottomrule
\end{tabular}
\end{table}

\section{Discussion}
\label{sec:discussion}

Our results establish TARA as a rigorous framework for quantum certification with distribution-free guarantees. The key findings are:

\begin{enumerate}
\item \textbf{CP validity for quantum data:} Theorem~\ref{thm:robustness} confirms that contextuality does not break conformal prediction, enabling its safe application to quantum systems.

\item \textbf{High detection accuracy:} TARA-$k$ achieves AUC = 0.96 against sophisticated LHV attacks, while TARA-$m$ provides anytime-valid streaming detection.

\item \textbf{Cross-platform robustness:} Validation on both IBM and IonQ demonstrates hardware-agnostic certification with 36\% security margins.

\item \textbf{Calibration leakage warning:} The 44-point inflation from same-distribution calibration suggests prior studies may have overestimated robustness. A comprehensive adversarial analysis of this vulnerability is provided in \cite{companion_p3}.
\end{enumerate}

\subsection{Related Work on Quantum Certification}

Several complementary approaches address quantum certification from different perspectives. Hardware validation of the Grothendieck constant \cite{companion_p2} has achieved measurements of $K_G = 1.408 \pm 0.006$ on trapped-ion systems (0.44\% deviation from $\sqrt{2}$) and established an empirical de Finetti relationship $\varepsilon = 0.498 \times S$ connecting CHSH violations to entanglement quantification. These hardware benchmarks provide independent validation of the detection regimes analyzed in the present work.

Adversarial analysis using generative adversarial networks \cite{companion_p3} has revealed fundamental detection limits: at mixing parameter $\alpha \geq 0.95$, classical adversaries can produce correlations that evade statistical detection methods entirely (AUC = 0.50). This work further analyzes the 44-point calibration leakage problem, demonstrating that same-distribution calibration is particularly vulnerable to adversarial exploitation and recommending cross-distribution protocols for robust certification.

\section{Conclusion}
\label{sec:conclusion}

We have introduced TARA (Test-by-Adaptive-Ranks), a comprehensive framework combining conformal prediction with martingale testing for quantum anomaly detection. Our work makes four principal contributions to the field of quantum certification.

First, we established theoretical foundations proving that conformal prediction validity is preserved for quantum data exhibiting contextuality. Theorem~\ref{thm:robustness} demonstrates that the sheaf-theoretic obstructions characterizing quantum nonlocality are cross-context phenomena that do not affect within-context exchangeability, ensuring that CP $p$-values remain uniformly distributed under the null hypothesis.

Second, we developed two complementary detection methods: TARA-$k$ for batch detection achieving ROC AUC = 0.96 against comprehensive LHV attacks, and TARA-$m$ for streaming detection with anytime-valid type-I error control via betting martingales. The TARA-$m$ approach enables real-time monitoring of quantum channels without pre-specifying sample sizes.

Third, cross-platform validation on IBM Torino (superconducting) and IonQ Forte Enterprise (trapped-ion) quantum processors demonstrated hardware-agnostic certification with 36\% security margins above the classical CHSH bound. Both architectures achieve similar maximum Bell violations, confirming the robustness of our detection methodology across fundamentally different physical implementations.

Fourth, we identified the 44-point calibration leakage problem, revealing that same-distribution calibration inflates detection performance from AUC = 0.50 to AUC = 0.94. This methodological concern may affect prior quantum certification studies and underscores the importance of cross-distribution calibration protocols. We recommend that future work report both same-distribution and cross-distribution metrics to quantify potential leakage.

These results establish TARA as a rigorous, distribution-free framework for practical quantum key distribution security, providing the statistical guarantees necessary for device-independent certification in near-term quantum devices.

\section*{Acknowledgements}

The author thanks IBM Quantum and Microsoft Azure Quantum for hardware access, and the open-source communities behind PyTorch, Qiskit, and scikit-learn.

\section*{AI Assistance Disclosure}

The author acknowledges the use of AI-assisted tools (OpenAI GPT, Anthropic Claude, Google Gemini) during manuscript preparation for literature review, code development assistance, and text refinement. The author takes full responsibility for all scientific content, has independently verified all experimental results, and confirms that all intellectual contributions and conclusions are the author's own work.

\section*{Data Availability and Reproducibility}

All code, experimental data, and analysis scripts required to reproduce the results in this paper are publicly available at \url{https://github.com/detasar/QCE}. The repository includes TARA-$k$ and TARA-$m$ detector implementations, IBM hardware measurement data (60,000 shots), and complete documentation enabling independent verification of all reported findings.

\section*{Competing Interests}

The author declares no competing interests.

\newpage
\section*{Supplementary Information}

Supplementary Information containing full proofs, extended experimental results, algorithm pseudocode, and code availability is provided as a separate document.

\end{document}


\maketitle

\tableofcontents
\newpage

\section{Theoretical Proofs}

\subsection{Proof of Theorem 1 (CP Robustness to Contextuality)}

\textbf{Theorem.} Conformal prediction with Mondrian conditioning provides valid coverage for quantum CHSH data.

\textbf{Proof.} Let $(X_i, Y_i, C_i)$ be feature-label-context triples where $C \in \{(0,0), (0,1), (1,0), (1,1)\}$ denotes the CHSH measurement setting.

1. Under Mondrian conformal prediction, we condition on each context $c$ separately:
$$p^M(X_{n+1}, y) = \frac{1 + \sum_{i: C_i = C_{n+1}} \mathbf{1}\{s_i \geq s(X_{n+1}, y)\}}{1 + |\{i : C_i = C_{n+1}\}|}$$

2. The key observation: CP validity depends only on \emph{exchangeability within each context}, not on the joint distribution across contexts.

3. Quantum contextuality, as characterized by sheaf-theoretic obstructions (Abramsky-Brandenburger), pertains to correlations \emph{across} contexts—the impossibility of a global section in the presheaf of distributions.

4. However, within each context $c$, quantum measurements yield exchangeable sequences: repeated measurements with the same settings produce i.i.d. outcomes.

5. Therefore, the calibration scores $\{s_i : C_i = c\}$ are exchangeable with the test score $s_{n+1}$, and the rank is uniformly distributed.

6. By the fundamental lemma of conformal prediction, the $p$-value is (super-)uniform.

\hfill $\square$

\subsection{Martingale Validity}

\textbf{Proposition.} Under the null hypothesis that data is generated by an LHV model, $\mathbb{E}[M_t] \leq 1$ for all $t$.

\textbf{Proof.} Under the null, conformal $p$-values are uniform. For the Kelly bet:
$$\mathbb{E}[1 + \beta(p)(p - 1/2)] = \mathbb{E}[1] = 1$$
since $\beta(p)(p-1/2)$ has expectation zero when $p \sim \text{Unif}(0,1)$. By the tower property, $\mathbb{E}[M_t] = 1$. \hfill $\square$

\section{Complete Algorithm Specifications}

\begin{algorithm}[H]
\caption{TARA-$k$ Detection Protocol}
\begin{algorithmic}[1]
\State \textbf{Input:} LHV calibration data $\mathcal{D}_{\text{LHV}}$, test data $D_{\text{test}}$, significance $\alpha$
\State \textbf{Phase 1: Feature Extraction}
\For{$D \in \mathcal{D}_{\text{LHV}} \cup \{D_{\text{test}}\}$}
    \State Compute CHSH $S = E_{00} + E_{01} + E_{10} - E_{11}$
    \State Compute click rates $p_A, p_B, p_{AB}, p_\emptyset$
    \State Compute TARA-$k = \sup_t |F_p(t) - t|$ (KS distance from uniform)\footnotemark
    \State Compute $\mathbb{E}|C_\alpha|$ (expected prediction set size)
    \State Form feature vector $\mathbf{f}(D) = [|S|, p_A, p_B, p_{AB}, p_\emptyset, \text{TARA-}k, \mathbb{E}|C_\alpha|]$
\EndFor
\State \textbf{Phase 2: Calibration}
\State Fit One-Class Elliptic Envelope on $\{\mathbf{f}(D)\}_{D \in \mathcal{D}_{\text{LHV}}}$
\State Estimate threshold $\tau_\alpha$ for desired FPR
\State \textbf{Phase 3: Detection}
\State Compute anomaly score $a = \text{OneClass}(\mathbf{f}_{\text{test}})$
\State Compute TARA-$k = \sup_t |\hat{F}_{\text{LHV}}(t) - \hat{F}_{\text{test}}(t)|$
\If{$a > \tau_\alpha$ \textbf{or} TARA-$k$ > KS$_\alpha$}
    \State \Return ``Quantum''
\Else
    \State \Return ``Classical''
\EndIf
\end{algorithmic}
\end{algorithm}

\footnotetext{The TARA-$k$ statistic (Test-by-Adaptive-Ranks, batch version) was originally denoted CMI (Conformal Miscoverage Index) in early development. Both names refer to the same Kolmogorov-Smirnov distance from uniformity.}

\begin{algorithm}[H]
\caption{TARA-$m$ Streaming Detection}
\begin{algorithmic}[1]
\State \textbf{Input:} Stream of measurements, threshold $\tau = 1/\alpha$
\State Initialize $M_0 \gets 1$
\For{$t = 1, 2, \ldots$}
    \State Receive measurement outcome $(x_t, z_t, a_t, b_t)$
    \State Compute conformal $p$-value from LHV-calibrated model
    \State $\beta_t \gets \lambda \cdot \mathrm{sign}(p_t - 1/2)$ \Comment{Kelly betting}
    \State $M_t \gets M_{t-1} \cdot (1 + \beta_t(p_t - 1/2))$
    \If{$M_t \geq \tau$}
        \State \Return ``Quantum detected at time $t$''
    \EndIf
\EndFor
\end{algorithmic}
\end{algorithm}

\section{Extended Experimental Results}

\subsection{ROC Analysis}

\begin{table}[H]
\centering
\caption{TARA-$k$ detection performance across feature configurations}
\begin{tabular}{lccc}
\toprule
Feature Set & ROC AUC & TPR@5\%FPR & TPR@1\%FPR \\
\midrule
Full ($|S|$ + CP) & 0.96 & 0.92 & 0.85 \\
CP-only (no $|S|$) & 0.83 & 0.72 & 0.58 \\
$|S|$-only & 0.91 & 0.84 & 0.76 \\
Click rates only & 0.78 & 0.62 & 0.48 \\
\bottomrule
\end{tabular}
\end{table}

\subsection{Hardware Validation Details}

\begin{table}[H]
\centering
\caption{Complete hardware validation results}
\begin{tabular}{lcc}
\toprule
Metric & IBM Torino & IonQ Forte \\
\midrule
CHSH $S$ & $2.725 \pm 0.042$ & $2.716 \pm 0.042$ \\
$E_{00}$ & 0.673 & 0.724 \\
$E_{01}$ & 0.671 & 0.704 \\
$E_{10}$ & 0.675 & 0.648 \\
$E_{11}$ & $-$0.672 & $-$0.640 \\
Classical margin & +36.3\% & +35.8\% \\
TARA-$k$ & 0.017 & 0.015 \\
TARA-$m$ log-wealth & 2.31 & 2.14 \\
Bell state fidelity & $\sim$0.90 & 0.984 \\
Detection decision & Quantum & Quantum \\
\bottomrule
\end{tabular}
\end{table}

\subsection{Calibration Leakage Analysis}

\begin{table}[H]
\centering
\caption{Impact of calibration protocol}
\begin{tabular}{lccc}
\toprule
Metric & Same-Dist & Cross-Dist & Inflation \\
\midrule
ROC AUC & 0.94 & 0.499 & +44 pp \\
Cohen's $d$ & 2.59 & $-$0.06 & Massive \\
TPR@5\%FPR & 92\% & 5\% & +87 pp \\
\bottomrule
\end{tabular}
\end{table}

\section{Code Availability and Reproducibility}

All experimental code, data, and analysis scripts required to reproduce the results are publicly available at:

\begin{center}
\url{https://github.com/detasar/QCE}
\end{center}

\begin{table}[H]
\centering
\caption{Repository structure for P1\_TARA\_Conformal\_Quantum}
\begin{tabular}{ll}
\toprule
File & Description \\
\midrule
\texttt{experiments/tara\_detectors.py} & TARA-$k$ and TARA-$m$ implementations \\
\texttt{experiments/tara\_batch\_detection.py} & Batch detection experiments \\
\texttt{experiments/data\_utils.py} & Data loading and CHSH computation \\
\texttt{data/ibm\_hardware\_real.csv} & IBM quantum hardware data (60,000 shots) \\
\bottomrule
\end{tabular}
\end{table}

\noindent The repository enables complete reproduction of all results presented in this paper.